\documentclass[onecollarge,natbib]{svjour2}
\bibpunct{[}{]}{;}{n}{}{,} 
\smartqed  
\journalname{To appear in Few Body Systems 55 (2014)}
\usepackage{amsmath}
\usepackage{amssymb}
\usepackage{graphicx,psfrag,epsfig}
\usepackage{afterpage}
\usepackage{mathrsfs,simplewick}

\DeclareMathOperator{\tr}{Tr}
\newcommand{\be}{\begin{equation}} \newcommand{\ee}{\end{equation}}
\newcommand{\ba}{\begin{eqnarray}} \newcommand{\ea}{\end{eqnarray}}
\newcommand{\bea}{\begin{eqnarray}} \newcommand{\eea}{\end{eqnarray}}
\newcommand{\bean}{\begin{eqnarray*}} \newcommand{\eean}{\end{eqnarray*}}

\newcommand{\s}[1]{{\scriptscriptstyle #1}}
\newcommand{\st}{{\scriptscriptstyle T}}
\def\slash#1{\setbox0=\hbox{$#1$}               
        \dimen0=\wd0                            
        \setbox1=\hbox{/} \dimen1=\wd1          
        \ifdim\dimen0>\dimen1                   
        \rlap{\hbox to \dimen0{\hfil/\hfil}}    
        #1                                      
        \else                                   
        \rlap{\hbox to \dimen1{\hfil$#1$\hfil}} 
        /                                       
        \fi}                                    %

\begin{document}

\title{Wilson Lines off the Light-cone in TMD PDFs 
}


\author{P.J. Mulders and M.G.A. Buffing       
}


\institute{P.J.~Mulders and M.G.A.~Buffing \at
              Theory Group, Nikhef \& 
              Department of Physics, Faculty of Science, VU University \\
              Amsterdam, the Netherlands\\
              \email{mulders@few.vu.nl} \\        
              \email{m.g.a.buffing@vu.nl}           
}

\date{Invited talk by P.J. Mulders at Lightcone 2013, 20-24 May 2013, Skiathos, Greece}

\maketitle

\begin{abstract}
Transverse Momentum Dependent (TMD) parton distribution functions (PDFs) also take into account the transverse momentum ($p_\st$) of the partons. The $p_\st$-integrated analogues can be linked directly to quark and gluon matrix elements using the operator product expansion in QCD, involving operators of definite twist. TMDs also involve operators of higher twist, which are not suppressed by powers of the hard scale, however. Taking into account gauge links that no longer are along the light-cone, one finds that new distribution functions arise. They appear at leading order in the description of azimuthal asymmetries in high-energy scattering processes. In analogy to the collinear operator expansion, we define a universal set of TMDs of definite rank and point out the importance for phenomenology.

\end{abstract}

\section{Introduction}
\label{intro}
In this contribution, we discuss the color-gauge-invariant definitions of transverse momentum 
dependent parton
distribution functions as they appear in azimuthal asymmetries in high-energy scattering 
processes. For this we use the formalism in which the parton distribution functions
are written as Fourier transform of nonlocal operator combinations of quark and gluon
operators. In order to be color-gauge-invariant, one needs the inclusion of gauge
links or Wilson lines, bridging the nonlocality. For the usual collinear parton
distribution functions depending on just one component of the parton momentum, 
namely the momentum fraction $x$, the gauge link dependence is easy to handle.
This changes when also transverse partonic momenta are included. The nonlocality
is no longer along a light-like direction and one can have different paths.

In order to introduce the basic notions, we first discuss the nonlocal operator
structure for parton distribution functions, closely following 
Ref.~\cite{Buffing:2011mj}. The starting point is a hard subprocess, such as
a two-to-two process with a truncated amplitude 
$\mathscr M (p_1,p_2;k_1,k_2)$, from which the wave functions of
the partons (Dirac spinors $u(p_1)$ for quarks, or polarizations 
$\epsilon(p_1)$ for gluons), are omitted. Rather than with plane 
wave spinors, the external partons are accounted for through
quark or gluon correlation or spectral functions, which are built 
from matrix elements of the form $\langle X\vert \psi(\xi)\vert P\rangle$
involving hadron states $\vert P\rangle$ replacing the free parton wave
function $\langle 0\vert\psi(\xi)\vert p\rangle$. An immediate complication
is the need to also include multi-parton matrix elements with the same 
states, such as $\langle X\vert A^\mu(\eta)\,\psi(\xi)\vert P\rangle$.

\begin{figure}
\begin{center}
\epsfig{file=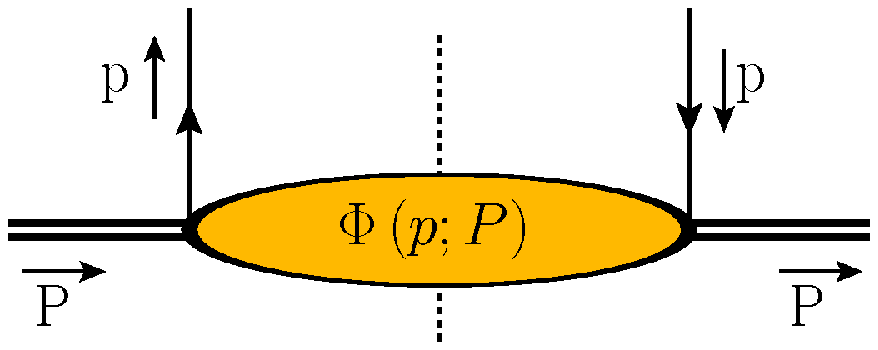,width=0.27\textwidth}
\hspace{1.5cm}
\epsfig{file=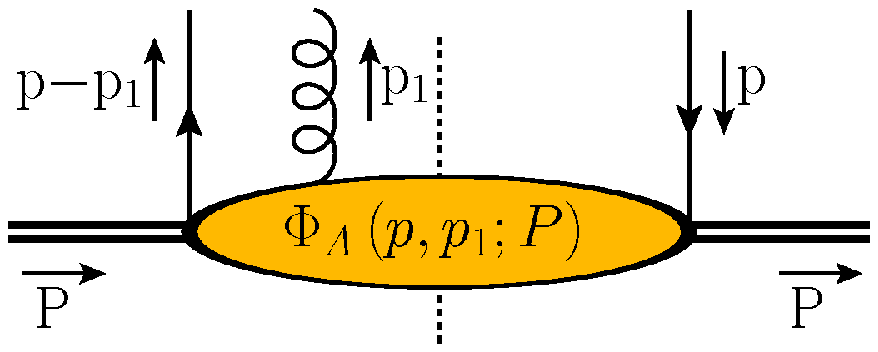,width=0.27\textwidth}
\hspace{1.5cm}
\epsfig{file=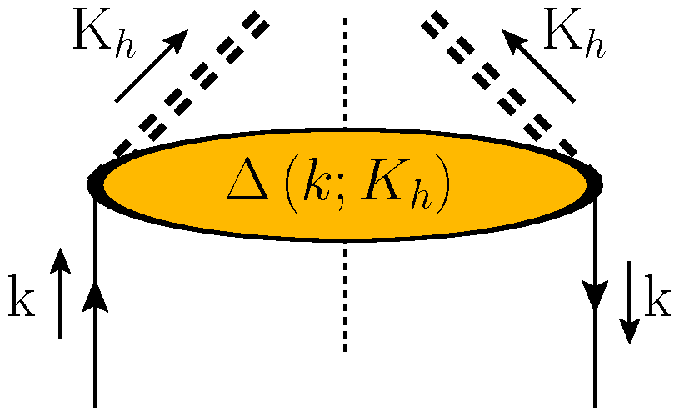,width=0.21\textwidth}
\\[0.2cm]
(a)\hspace{6.0cm} (b)\hspace{5.5cm} (c)
\end{center}
\caption{\label{fig1}
The pictorial momentum space representation of the quark-quark correlator (a) and the
quark-quark-gluon correlator for 
distribution functions (b) and a quark-quark correlator for
fragmentation functions (c).} 
\end{figure}

At high energies, the matrix elements appear as squared contributions in the 
correlators (including Dirac space indices $i$ and $j$),
\bea
\Phi_{ij}(p;P) & = &
\sum_X\int \frac{d^3P_X}{(2\pi)^3\,2E_X}
\ \langle P\vert \overline\psi_j(0)\vert X\rangle
\,\langle X\vert \psi_i(0)\vert P\rangle\,\delta^4(p+P_X-P)
\nonumber \\ & = &
\frac{1}{(2\pi)^4}\int d^4\xi\ e^{i\,p\cdot \xi}
\ \langle P\vert \overline\psi_j(0)\,\psi_i(\xi)\vert P\rangle ,
\eea
pictorially represented in Fig.~\ref{fig1}(a). 
Usually, a summation over color indices is understood. This means that
we will have $\Phi(p) = {\rm Tr}_c\bigl[\Phi(p)\bigr]$, where
$\Phi_{ij}(p)$ is considered to be also a matrix in color space,
made explicit 
$\Phi_{ij;rs} \propto \psi_{ir}(\xi)\,\overline\psi_{js}(0)$.
Including gluon fields one has quark-quark-gluon correlators like
\be
\Phi^\mu_{A\,ij}(p,p_1;P) =
\frac{1}{(2\pi)^8}\int d^4\xi\,d^4\eta
\ e^{i\,(p-p_1)\cdot \xi}\ e^{i\,p_1\cdot \eta}
\ \langle P\vert \overline\psi_j(0)\,A^\mu(\eta)\,\psi_i(\xi)\vert P\rangle,
\label{quarkgluonquark}
\ee
illustrated in Fig.~\ref{fig1}(b), 
and similarly matrix elements with more partons.
The color structure of the field combination $\psi_r(\xi)\overline\psi_s(0)$ 
in the quark-quark-gluon correlator now actually has a color octet structure,
with the simplest color trace being ${\rm Tr}_c[\psi\overline\psi A^\mu]$ if 
we write $A^\mu = A^{\mu a}T^a$ as a matrix-valued field.
The corresponding correlators describing fragmentation into hadrons
is for quarks given by
\begin{eqnarray}
\Delta_{ij}(k;K_h) & = & \sum_X \frac{1}{(2\pi)^4}
\int d^4\xi\ e^{-ik\cdot \xi}\,
\langle 0 \vert \psi_i(0) \vert K_h, X \rangle
\langle K_h,X \vert \overline \psi_j(\xi)
\vert 0 \rangle \nonumber \\
& = & \frac{1}{(2\pi)^4}\int d^4\xi\ e^{-ik\cdot \xi}\,
\langle 0 \vert \psi_i (0) a_h^\dagger
a_h \overline \psi_j(\xi) \vert 0 \rangle,
\label{frag}
\end{eqnarray}
pictorially represented by the blob in Fig.~\ref{fig1}(c).
An averaging over color indices is implicit,
thus $\Delta(k) = \frac{1}{N_c}{\rm Tr}_c\bigl[\Delta(k)\bigr]$ with
again $\Delta(k)$ a diagonal matrix in color space. 
The second expression in the above involves hadronic creation and annihilation 
operators $a_h^\dagger\vert 0\rangle = \vert K_h\rangle$.

\begin{figure}
\begin{center}
\epsfig{file=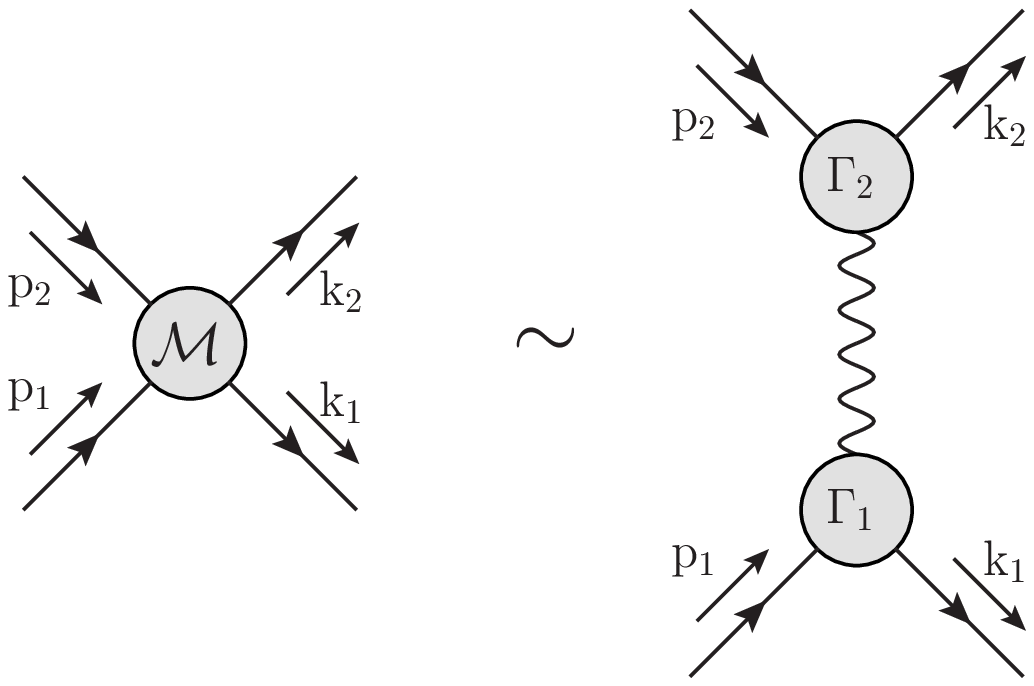,width=0.34\textwidth}
\hspace{2cm}
\epsfig{file=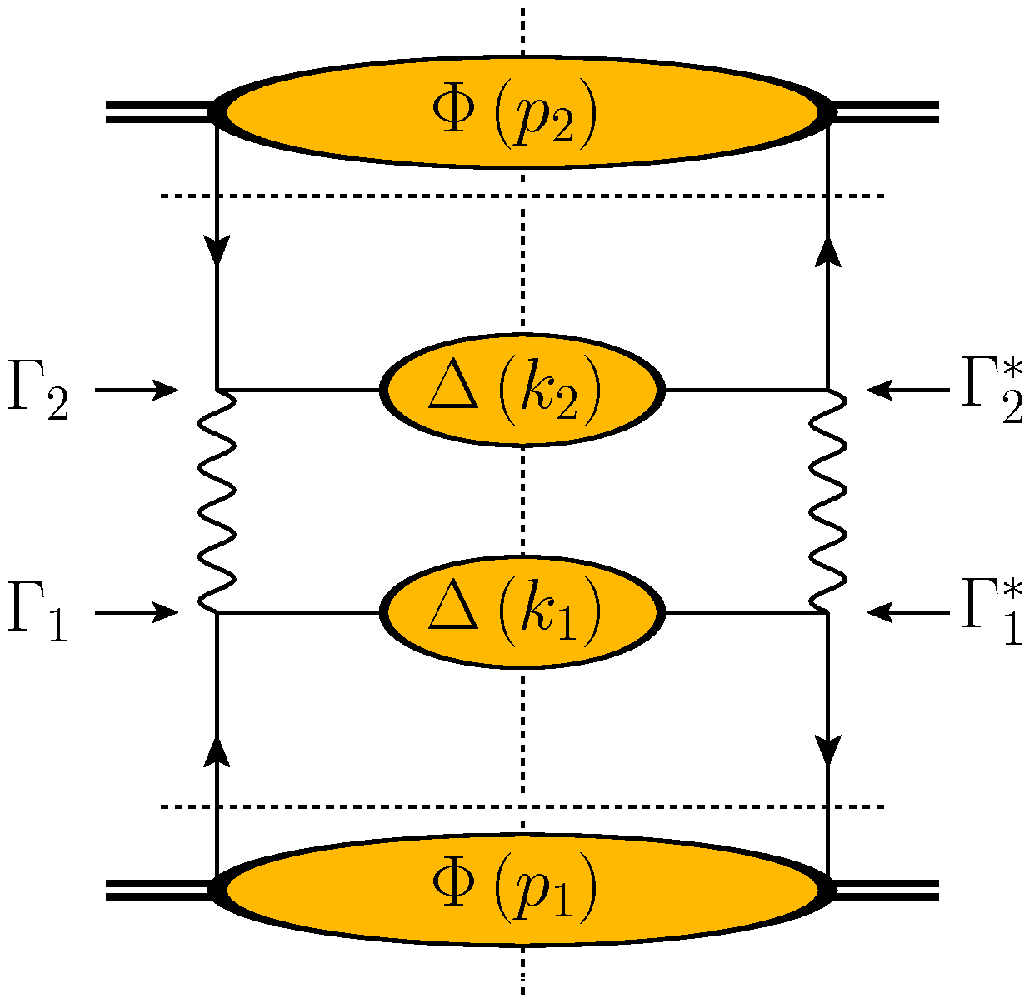,width=0.3\textwidth}
\\[0.2cm]
\mbox{}\hspace{1cm}(a)\hspace{6.5cm} (b)
\end{center}
\caption{\label{fig2}
To illustrate the inclusion of correlators, we use the
hard amplitude with one particular color flow for the quark lines as
shown in (a). The squared amplitude needed for the cross section 
of the scattering process initiated by two hadrons is shown in (b). }
\end{figure}
The inclusion of the correlators in the description of a scattering 
process is similar to the inclusion of quark spinors or gluon polarizations.
In the expression for the cross section of the (semi)-inclusive
process $H_1(P_1) + H_2(P_2) \rightarrow h_1(K_1) + h_2(K_2) + \dots$
partonic momenta are approximately collinear, $P_1{\cdot}p_1 \sim P_2{\cdot}p_2 
\sim K_1{\cdot}k_1 \sim K_2{\cdot}k_2$ being of a hadronic mass scale, 
$M^2 \sim 1$ GeV$^2$. This is to be compared to the usual hard invariants 
in the full or the partonic process such as
$s \approx 2\,P_1{\cdot}P_2$, $t_1 \approx -2\,K_1{\cdot}P_1$,
$\hat s \approx 2\,p_1{\cdot}p_2$, $\hat t \approx -2\,k_1{\cdot}p_1$
(we will refer to this scale as the squared hard scale, $Q^2 \gg M^2$).

The squared partonic amplitude $\vert \mathscr M\vert^2$ is convoluted 
with the correlators $\Phi(p,P)$ and $\Delta(k,K_h)$. As illustrated in
Figs~\ref{fig2}(a) and (b) for an example in which (for simplicity) no color is
exchanged in the hard process, the cross section is of the form
\be
d\sigma \sim 
{\rm Tr}_c\bigl[\Phi(p_1)\,\Gamma_1^\ast\,\Delta(k_1)\,\Gamma_1\bigr]
\,{\rm Tr}_c\bigl[\Phi(p_2)\,\Gamma_2^\ast\,\Delta(k_2)\,\Gamma_2\bigr],
\label{facto1}
\ee
where ${\rm Tr}_c\bigl[\ldots\bigr]$ parts are traced over color.
In the case that the vertices $\Gamma$ don't have any color structure,
one can, because of the simple color singlet structure of $\Phi$ and $\Delta$ 
in the quark-quark correlators, perform the color trace separately 
for $\Phi$ and $\Delta$,
${\rm Tr}_c[\Phi(p)\,\Gamma^\ast\,\Delta(k)\,\Gamma] 
= {\rm Tr}_c[\Phi(p)]\tfrac{1}{N_c}{\rm Tr_c}[\Delta(k)]
\,\Gamma\,\Gamma^\ast$
(one summed and one averaged) and the cross section can be written in 
terms of the color-traced entities 
\be
d\sigma \sim 
\Phi(p_1)\,\Phi(p_2)
\,\underbrace{\Gamma_1\,\Gamma_1^\ast\,\Gamma_2\,\Gamma_2^\ast}_{\hat\Sigma}
\,\Delta(k_1)\,\Delta(k_2),
\label{facto1a}
\ee
where the remaining contractions are Dirac space and Lorentz indices, which
have been suppressed in both Eqs~\ref{facto1} and \ref{facto1a}.
This expression still needs to be integrated over the parton momenta,
which will be discussed next and of course should be extended with all possible
correlators containing quark and gluon fields, in which cases color
traces become more complicated.

The restriction to hard kinematics limits the number 
of diagrammatic contributions, although even at leading order, there 
still are many gluon contributions as will be discussed in 
Section~\ref{section2}. 
For parton momenta relevant in a hadron correlator
we use the Sudakov decomposition,
\be
p = x\,P + p_\st + \underbrace{(p{\cdot}P - x\,M^2)}_{\sigma}\,n,
\label{sudakov}
\ee
where $n$ is a light-like vector $n$, satisfying
$P{\cdot}n = 1$ which can come from another
hard (external) momenta, e.g.\ $n = P^\prime/P^\prime{\cdot}P$.
The momentum fraction $x = p{\cdot}n = p^n$ is $\mathscr O(1)$. For
any contractions with vectors outside the correlator $\Phi(p,P)$ one
has $P \sim Q$, $p_\st \sim M$ and $n \sim 1/Q$.
Note that if $n$ is an exact light-like vector, 
one can construct {\em two} exact conjugate null-vectors,
$n_+ = P -\tfrac{1}{2}\,M^2\,n$ and $n_- = n$,
satisfying
$n_+{\cdot}n_- = 1$ and $n_+^2 = n_-^2 = 0$, that can be used to define
light-cone components
$a^\pm = a{\cdot}n_\mp$ (thus $x = p{\cdot}n_- = p^+$).
Symmetric and antisymmetric `transverse' projectors are defined as
$g_\st^{\mu\nu} = g^{\mu\nu} - n_+^{\{\mu}n_-^{\nu\}}$ and 
$\epsilon_\st^{\mu\nu} = \epsilon^{n_+n_-\mu\nu}
= \epsilon^{-+\mu\nu}
= \epsilon^{Pn\mu\nu}$.
In view of the relative importance of the components in this integration,
one can, upon neglecting any $M^2/Q^2$ contributions in the cross section,
integrate within a soft correlator over $p{\cdot}P$ (i.e.\ $p^-$) to 
obtain the TMD correlator
\be
\Phi(x,p_\st;n) = \int d\,p{\cdot}P\ \Phi(p;P)
= \left. \int \frac{d\,\xi{\cdot}P\,d^2\xi_\st}{(2\pi)^3}\ e^{i\,p\cdot \xi}
\ \langle P\vert \overline\psi(0)\,\psi(\xi)\vert P\rangle \right|_{LF}\ ,
\label{phi-tmd}
\ee
which we will still consider as the unintegrated correlator.
On the left-hand side the dependence on the hadron momentum $P$ has been
suppressed.
In the TMD correlator the nonlocality is restricted to the 
light-front (LF: $\xi{\cdot}n = \xi^+ = 0$) and the correlator depends 
on $x = p{\cdot}n$ and $p_\st$. This {\em light-front} correlator is actually
at equal (light-cone) time and time-ordering, thus, is automatic.
This allows a direct interpretation of the correlator as a forward antiparton-hadron
scattering amplitude, i.e.\ a Green function, untruncated in the parton 
legs~\cite{Jaffe:1983hp}. This is the
case for both collinear and TMD correlators~\cite{Diehl:1998sm}. This 
identification has been very important in deep inelastic 
processes~\cite{Landshoff:1971xb}, allowing the
use of analyticity and unitarity properties of field theories, at least 
under the assumption that these properties apply to QCD. We will also need 
this later for fragmentation correlators.

Finally, the {\em light-cone} correlators are the collinear correlators containing the
parton distribution functions depending only on the light-cone momentum
fraction $x$, obtained upon integration over both $p{\cdot}P$ and $p_\st$,
\be
\Phi(x;n) = \int d\,p{\cdot}P\,d^2p_\st\ \Phi(p;P)
= \left. \int \frac{d\,\xi{\cdot}P}{(2\pi)} \ e^{i\,p\cdot \xi}
\ \langle P\vert \overline\psi(0)\,\psi(\xi)\vert P\rangle \right|_{LC}\ ,
\label{phi-coll}
\ee
where the subscript LC refers to light-cone, implying $\xi{\cdot}n$ = $\xi_\st$
= 0. This integration is generally allowed in hard 
processes up to $M^2/Q^2$ contributions and also up to contributions coming 
from the tails, e.g.\ logarithmically divergent contributions proportional to
$\alpha_s(p_\st^2)/p_\st^2$ tails~\cite{Bacchetta:2008xw}. 
The treatment of these in principle logarithmically divergent contributions 
require going beyond the tree-level resummation and consider next-to-leading 
order (NLO) QCD. 
In diagrammatic language these contributions involve ladder graphs describing
the emission of gluons into the final state, relevant for the evolution of 
the correlators.

We end this section with a note on the measurability of the transverse momentum dependent (TMD) correlators. The collinear correlators are relevant in hard processes 
in which only hard scales (large invariants $\sim Q^2$ or ratios thereof, angles, 
rapidities) are measured.
If one considers hadronic scale observables (transverse
momenta within jets or slightly off-collinear configurations) one will need
the TMD correlators for a full treatment. 
To identify the appropriate observable momentum for TMDs one must realize
that the calculation of the cross section as schematically indicated in
Eq.~\ref{facto1} involves momentum conservation at the partonic level,
for instance $p+q = k$ for semi-inclusive deep inelastic scattering 
(DIS with $q$ being the space-like virtual photon momentum, $q^2 = -Q^2$) or
$p_1 + p_2 = k_1 + k_2$ for a two-to-two process like the Drell-Yan process 
(DY where the final state is a lepton-pair with time-like momentum $q = k_1 + k_2$ 
and $q^2 = Q^2$).

Realizing that in the parametrization of partonic momenta, the momentum
conservation can be used to identify the momentum fractions at high energies 
with scaling variables, e.g. for SIDIS one has for the fractions in the Sudakov
expansions $p \approx xP + p_\st$ and $k = K/z + k_\st$
\be 
x \approx x_B \equiv \frac{Q^2}{2P{\cdot}q} \quad \mbox{and} \quad
z \approx z_h \equiv \frac{P{\cdot}K}{P{\cdot}q},
\ee
equating $x$ to the Bjorken scaling variable and $z$ to the appropriate ratio
of components of hadron momentum and photon momentum. Since these relations
are correct up to $1/Q^2$ corrections, one finds that in the transverse
directions one can identify
\be
q_\st \equiv q + x_B P - K/z_h = k_\st - p_\st, 
\ee
as the measurable transverse momentum up to order $1/Q$ corrections.
Similarly for hadron initiated processes~\cite{Ralston:1979ys,Tangerman:1994eh,Boer:1999mm,Pisano:2013cya} one can access 
transverse momenta. The best example is DY where one identifies the fractions with 
scaling variables
\be 
x_1 = \frac{q{\cdot}P_2}{P_1{\cdot}P_2} \quad \mbox{and} \quad
x_2 = \frac{q{\cdot}P_1}{P_2{\cdot}P_1}
\ee
and finds that the relevant measure for transverse momentum is
\be
q_\st \equiv q - x_1 P - x_2 P_2 = p_{1\st} + p_{2\st}. 
\ee
In both cases one is left with a convolution of the partonic transverse
momenta in the hadrons. One can extend this to off-collinearity of jets
or produced hadrons in the final state in more complicated hadron-hadron
scattering processes. 
 
\section{Color gauge invariance
\label{section2}}

The correlators encompass the information on the soft parts. They
depend on the hadron and quark momenta $P$ and $p$ (and in general also
spin vectors). Depending on the Lorentz and Dirac structure of the matrix 
elements involved one can look for the pieces in the correlator
that show up as the most dominant matrix elements among the 
contributions in the hard process. These are those that have the
maximum number of contractions with $n$, which minimizes the powers
of $M$ that after contraction of open indices inevitably is the scale
of the hadronic matrix elements. 
Including also gluon fields, they are
\be
\langle\ \overline\psi(0)\slash n\psi(\xi)\ \rangle
\qquad \mbox{and} \qquad
\langle\ G^{n\alpha}(0)G^{n\beta}(\xi)\ \rangle,
\label{nonlocalfields}
\ee
(the latter with transverse indices $\alpha$ and $\beta$). 
The two matrix elements above have canonical dimension two. 
The corresponding local matrix elements, 
$\overline\psi(0)\,\slash n\,\psi(0)$ and
$G^{n\alpha}(0)\,G^{n\beta}(0)$ for quarks and gluons, 
respectively, are color-gauge-invariant (twist two) operators.
The nonlocal combinations in Eq.~\ref{nonlocalfields}, however, are not gauge-invariant. Expanded into local operators, the expansion would involve 
operator combinations with derivatives such as 
$\overline\psi(0)\,\slash n\,\partial^n\ldots \partial^n\,\psi(0)$.
Color gauge invariance in the correlators requires in the
local matrix elements covariant derivatives or in the nonlocal
matrix elements the presence of a gauge link connecting the two fields. 
For the light-cone correlators the gauge link corresponds to
the inclusion of an arbitrary number of `leading' gluon fields 
$A^n(\eta)$ in the field combinations in Eq.~\ref{nonlocalfields} 
which are resummed into a gauge link
$\overline\psi(0)\,U^{[n]}_{[0,\xi]}\,\psi(\xi)$
= ${\rm Tr}_c\bigl[U^{[n]}_{[0,\xi]}\,\psi(\xi)\overline\psi(0)\bigr]$, 
given by
\be
U^{[n]}_{[0,\xi]}
= {\mathscr P}\,\exp\left(-i\int_0^\xi d\,\eta{\cdot}P\ n{\cdot}A(\eta)\right).
\label{glcol}
\ee
Including this gauge link, the nonlocal operator combinations 
\be
\left.\langle\ \overline\psi(0)\slash n\,U^{[n]}_{[0,\xi]}\psi(\xi)
\ \rangle\right|_{LC}
\qquad \mbox{and} \qquad
\left.\langle\ G^{n\alpha}(0)\,U^{[n]}_{[0,\xi]}
G^{n\beta}(\xi)\,U^{[n]}_{[\xi,0]}\ \rangle\right|_{LC}
\label{nonlocalfields-2}
\ee
can be expanded into twist two operators 
$\overline\psi(0)\,\slash n\,D^n\ldots D^n\,\psi(0)$ and
$G^{n\alpha}(0)\,D^n\ldots D^n\,G^{n\beta}(0)$ for quarks and gluons, 
respectively (number of $D^n$'s is the spin of these operators).
Also TMD correlators require a gauge link, but the separation of the
two fields is no longer a simple light-like one and they involve
derivatives with transverse indices. It is important to realize that
in principle any gauge link with an arbitrary path gives a 
gauge-invariant combination. What is the appropriate link contributing
at leading order (in $M/Q$) in a given
hard scattering process, however, is calculable~\cite{Bomhof:2006dp}.

\begin{figure}
\begin{center}
\epsfig{file=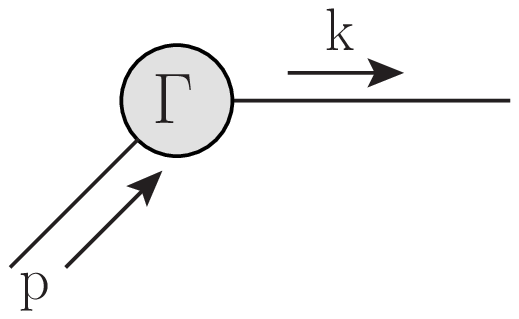,width=0.15\textwidth}
\hspace{1cm}
\epsfig{file=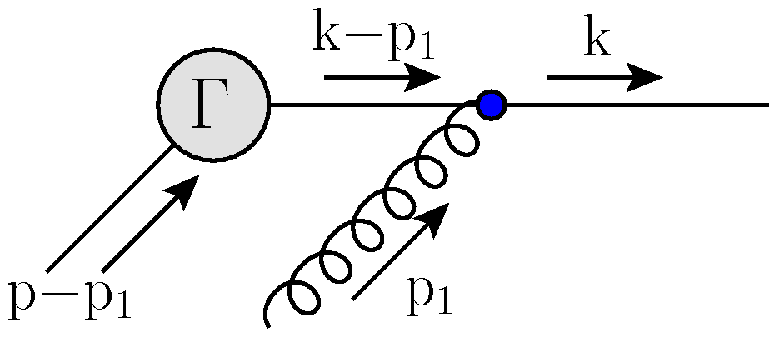,width=0.23\textwidth}
\hspace{1cm}
\epsfig{file=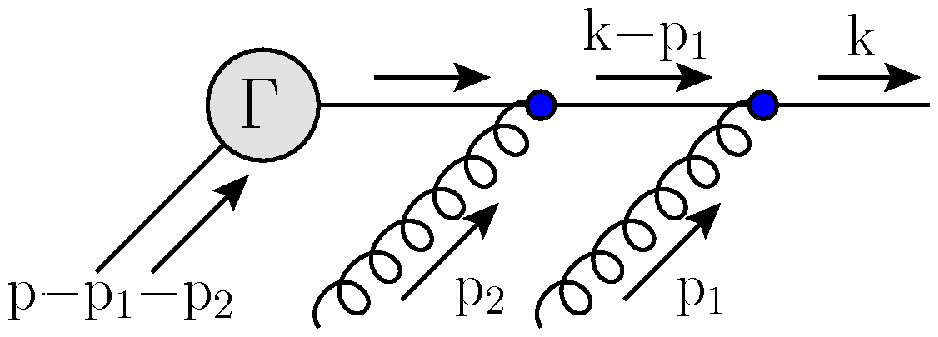,width=0.27\textwidth}
\\[0.2cm]
$A_0$\hspace{5.5cm} $A_1$\hspace{5cm} $A_2$
\end{center}
\caption{\label{fig3}
Inclusion of collinear gluons from 
$\Phi_{A\ldots A}(p-p_1\ldots - p_N,p_1,\ldots ,p_N)$ 
coupling to an outgoing (colored) quark line with momentum $k$.
These insertions give rise to the gauge link $U_+^{[k]}[p]$ in Eq.~\ref{facto2}.}
\end{figure}

The calculation involves diagrams with additional gluon fields in the 
correlator. The important ones at leading order are the $A^n$ gluons
which do not increase the canonical dimension and hence also appear
at leading order. Couplings of these gluons into the hard part cancel
using Ward identities and the couplings to the external lines (see
Fig.~\ref{fig3}) produce
a Wilson line running from the positions $0$ or $\xi$ in the parton fields
to $\xi^- = \pm \infty$ depending on the external line being an incoming
parton or outgoing parton.
\begin{figure}
\begin{center}
(a)\hspace{0.5cm}\includegraphics[width=4.5cm]{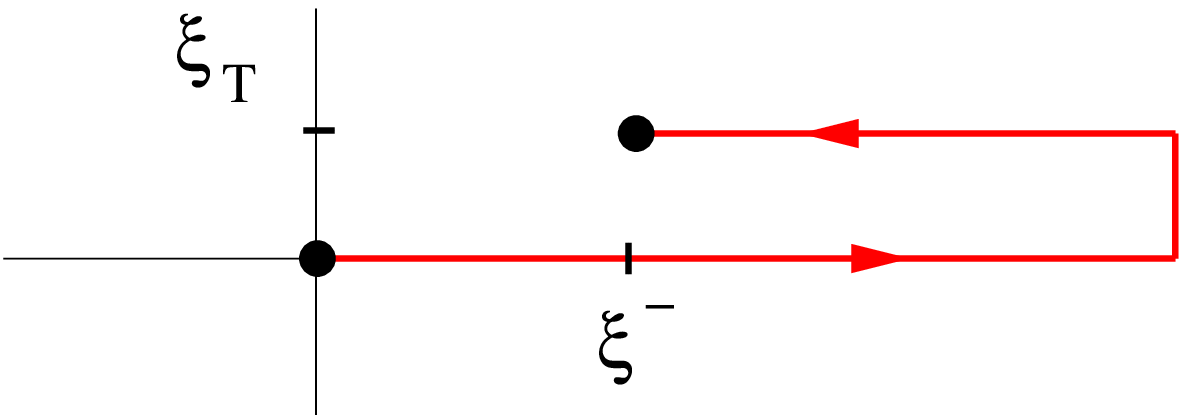}
\hspace{1cm}
(b)\hspace{0.5cm}\includegraphics[width=4.5cm]{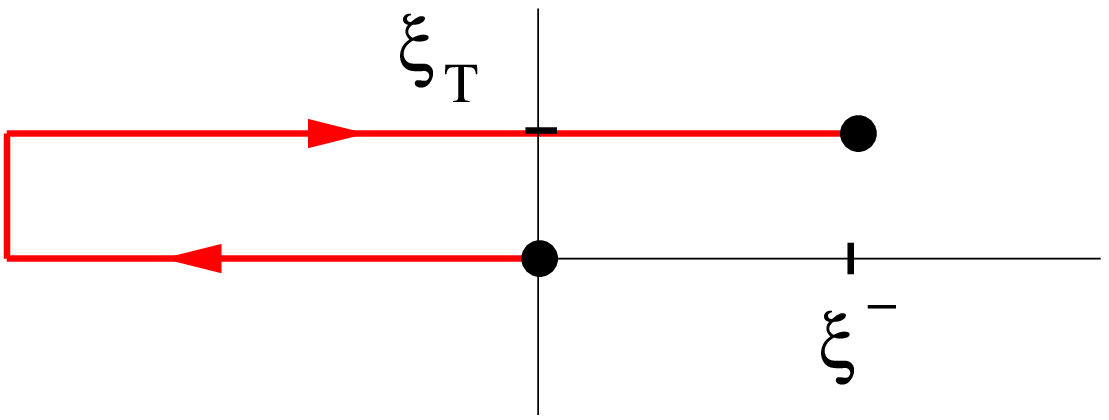}
\end{center}
\caption{The staple-like gauge link structure in the quark-quark
correlator $\Phi$ in SIDIS (a) and DY (b) respectively. 
\label{simplelinks}}
\end{figure}
We note that the gauge link also involves transverse gluons
showing that in processes involving more hadrons the effects of
transverse gluons are not necessarily suppressed. 
Integration over transverse momenta implies in the correlator, which is
a Fourier transform, that $\xi_\st = 0_\st$, in which case the
$U^{[+]}$ and $U^{[-]}$ links reduce to a unique collinear link
connecting $0$ and $\xi^-$.
Technically, the transverse gluons emerge as boundary terms at light-cone 
infinity~\cite{Belitsky:2002sm,Boer:2003cm} that are needed to rewrite
transverse gluon fields $A^\alpha$ into field strengths $G^{n\alpha}$.
Physically they can also be seen to correspond to soft gluons as
shown in explicit model calculations~\cite{Brodsky:2002cx,Brodsky:2002rv}. 
If only one hadron appears in the high-energy scattering process, this
produces the color-gauge-invariant matrix elements with for transverse
momentum dependent correlators (cf.\ Eq.~\ref{phi-tmd}) as simplest
gauge links the ones shown in Fig.~\ref{simplelinks} for quark correlators 
and in Fig.~\ref{simplelinks2} for gluon correlators. For collinear
correlators (cf.\ Eq.~\ref{phi-coll}) the staple-like gauge links
reduce to a unique straight-line gauge link with single or double
color lines for quarks or gluons respectively. 
 
\begin{figure}
\begin{center}
\epsfig{file=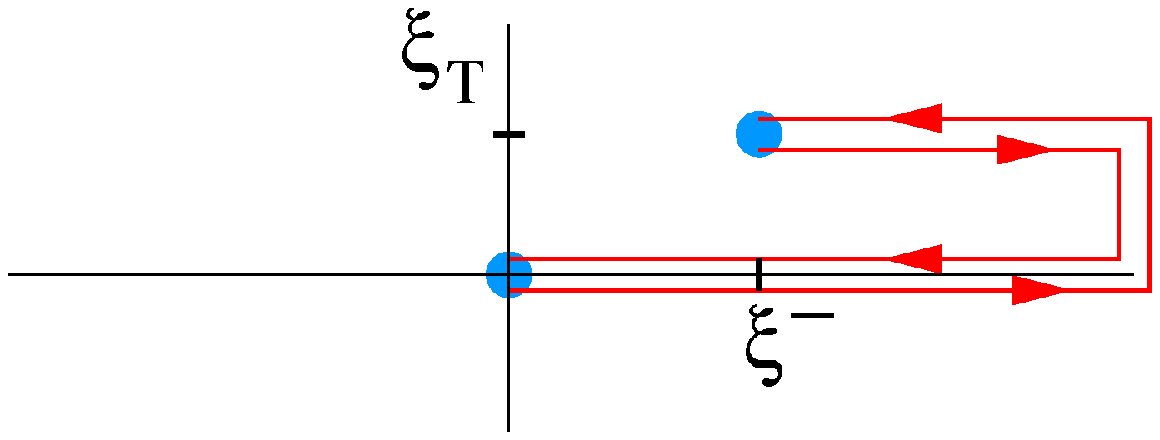,width=5.0cm}
\hspace{0.5cm}
\epsfig{file=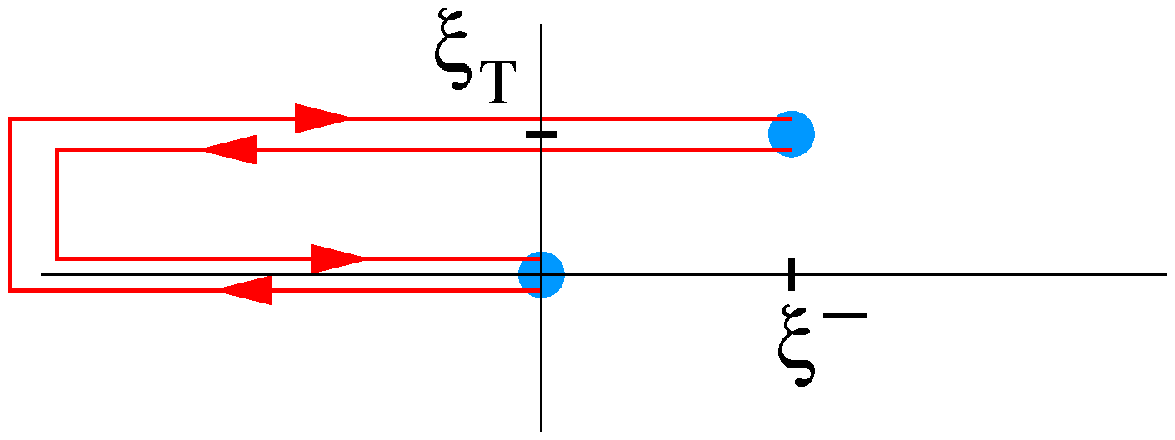,width=5.0cm}
\\
(a) \hspace{6cm} (b)
\\
\epsfig{file=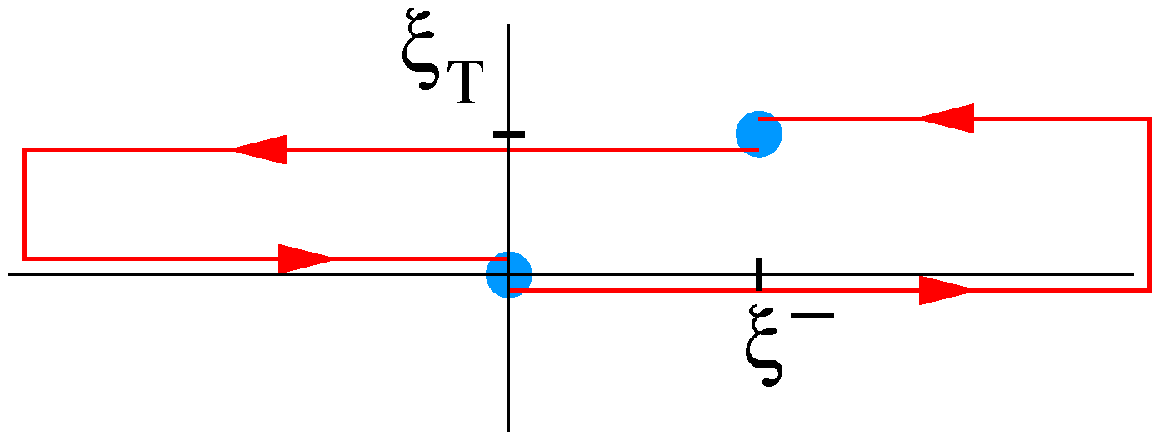,width=5.0cm}
\hspace{0.5cm}
\epsfig{file=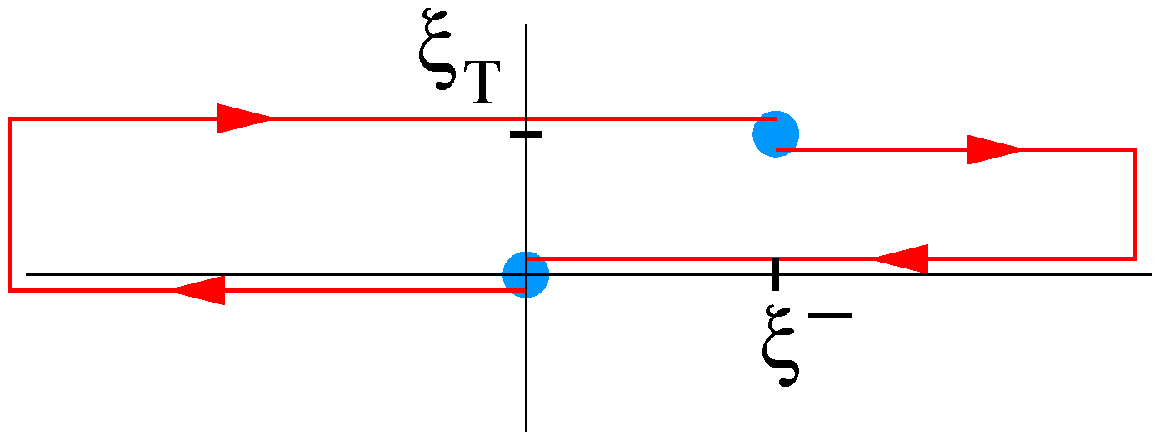,width=5.0cm}
\\
(c) \hspace{6cm} (d)
\end{center}
\caption{\label{simplelinks2} The gauge links for gluon TMDs. We note that
for gluons the double color line may `split' in the hard part and connect
to an initial state {\em and} a final state parton giving rise to the 
gauge link structures in (c) and (d).}
\end{figure}

\section{Color entanglement}

The color structure of various
correlators becomes entangled if the color flow is more 
complicated~\cite{Bomhof:2006dp,Bomhof:2004aw,Bacchetta:2005rm,Bomhof:2006ra,Bomhof:2007xt,Rogers:2010dm}. Following again Ref.~\cite{Buffing:2011mj} we
find for the gluon insertions coming from a particular correlator and
coupling to an outgoing fermion line a Wilson line connecting 
to light-cone $+\infty$. E.g.\ the gauge link $U_+^{[k_1]}[p_1]$ emerges 
from diagrams as shown in Fig.~\ref{fig3}.
Including all multi-gluon interactions originating from $\Phi(p_1)$ in
Fig.~\ref{fig2}(b) and transverse pieces, we get the result
\bea
d\sigma & \sim & {\rm Tr}_c\bigl[\Phi(p_1)\,\Gamma_1^\ast
\,U_+^{[k_1]\dagger}[p_1]\Delta(k_1)U_+^{[k_1]}[p_1]\,\Gamma_1\bigr]
\nonumber \\&& \mbox{}\times
{\rm Tr}_c\bigl[U_-^{[p_2]\dagger}[p_1]\,\Phi(p_2)\,U_-^{[p_2]}[p_1]
\,\Gamma_2^\ast
\,U_+^{[k_2]\dagger}[p_1]\,\Delta(k_2)\,U_+^{[k_2]}[p_1]\,\Gamma_2\bigr],
\label{facto2}
\eea
in which the (color charge of the) Wilson line is stuck in the color traces
at the `positions' corresponding to the external parton lines. Although the 
light-like directions in the gauge links involve different light-like directions,
these are all `orthogonal' light-like directions to $p_1$ and can at leading order
simply be replaced by a single generic null-vector $n$.

\begin{figure}[b]
\begin{center}
\epsfig{file=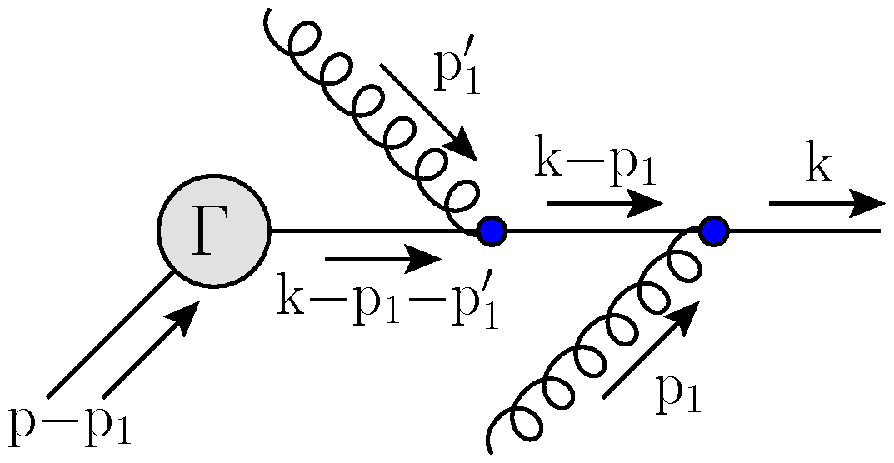,width=0.22\textwidth}
\hspace{1cm}
\epsfig{file=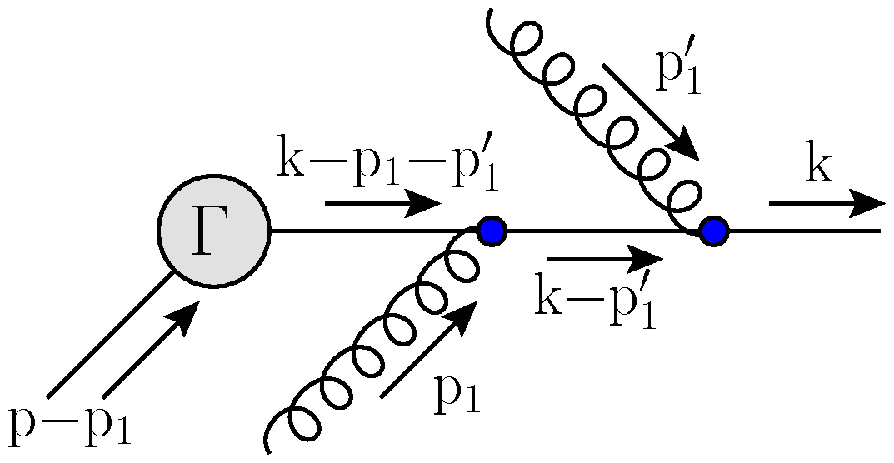,width=0.22\textwidth}
\hspace{1cm}
\epsfig{file=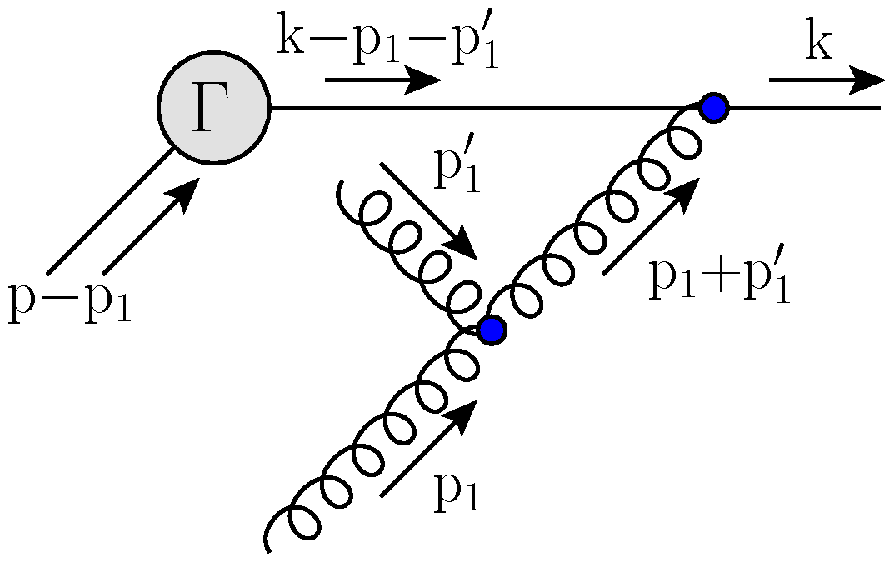,width=0.22\textwidth}
\\[0.2cm]
(a)\hspace{5.0cm} (b)\hspace{5.0cm} (c)
\end{center}
\caption{\label{figB1}
The gluon insertions on an outgoing quark line coming from two different
soft pieces, one from $\Phi(p)$ and one from $\Phi(p^\prime)$, respectively.}
\end{figure}
\begin{figure}
\begin{center}
\epsfig{file=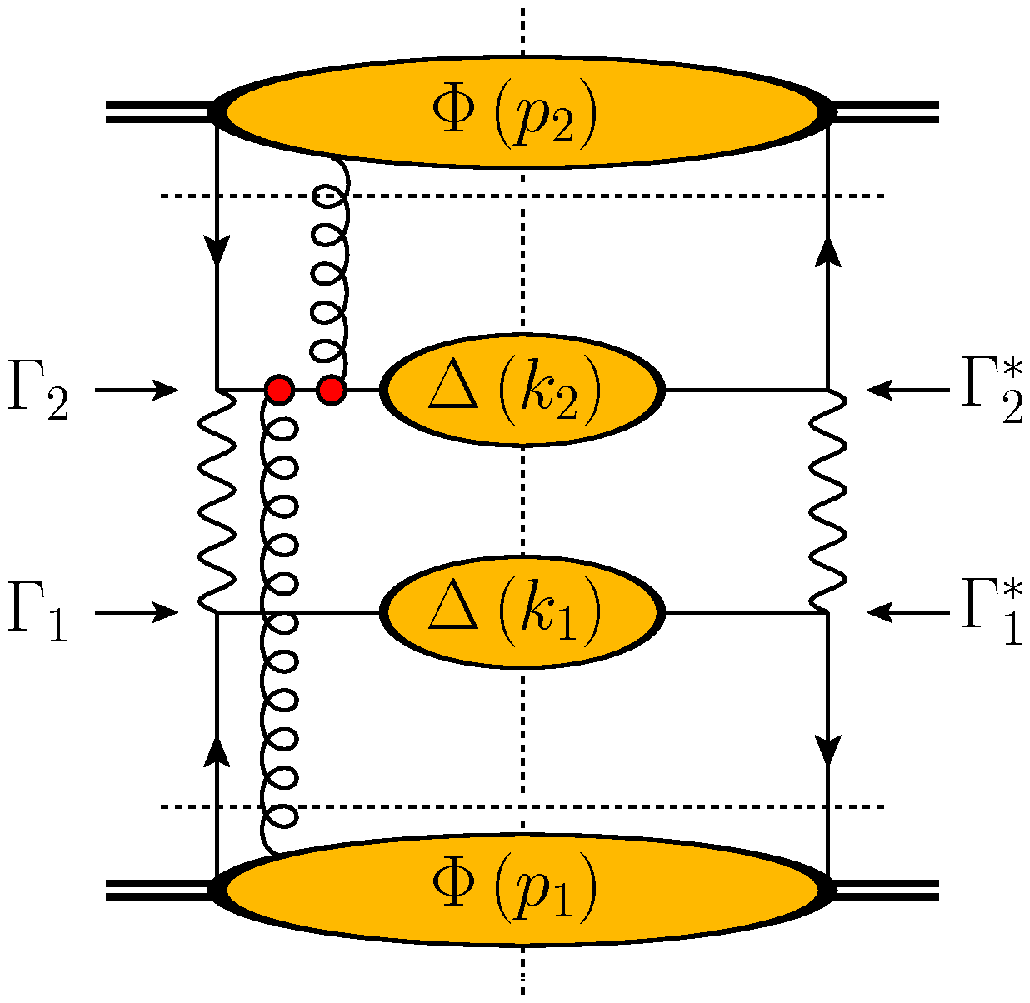,width=0.3\textwidth}
\hspace{0.5cm}
\epsfig{file=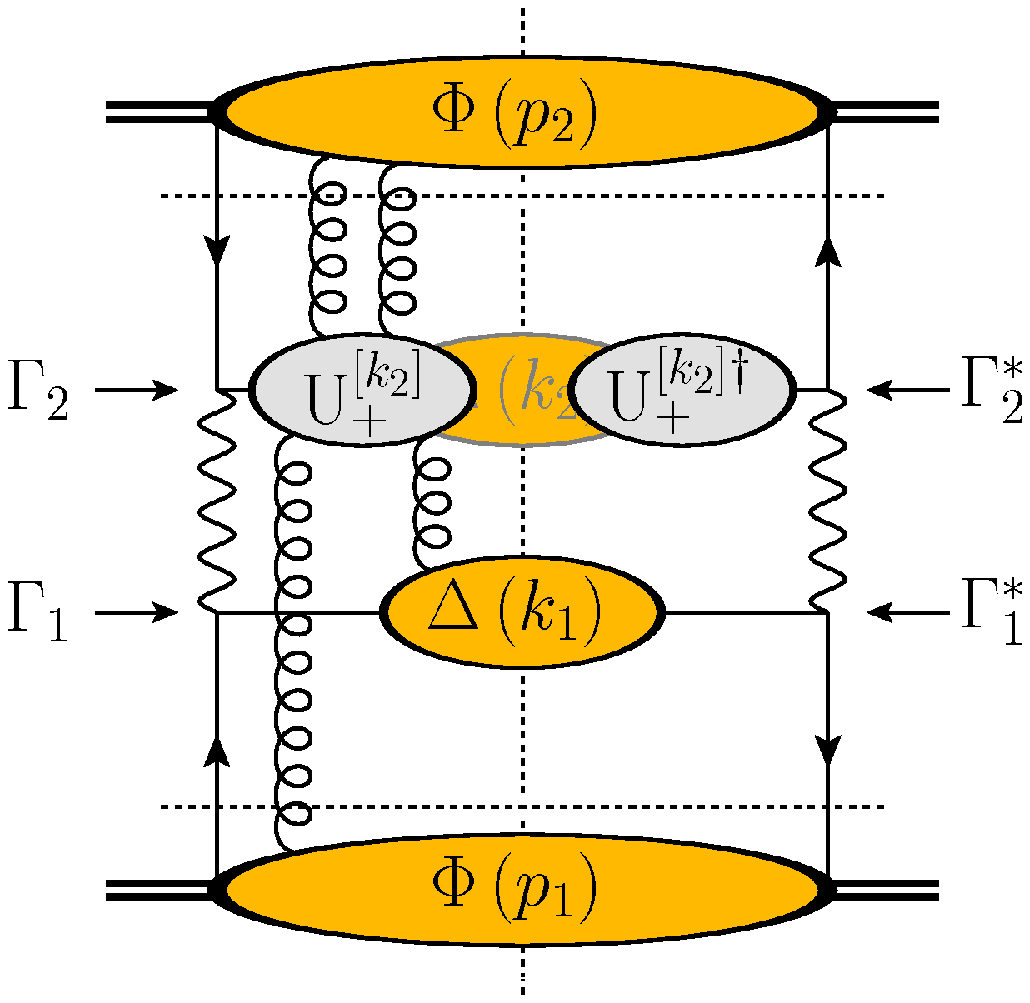,width=0.3\textwidth}
\hspace{0.5cm}
\epsfig{file=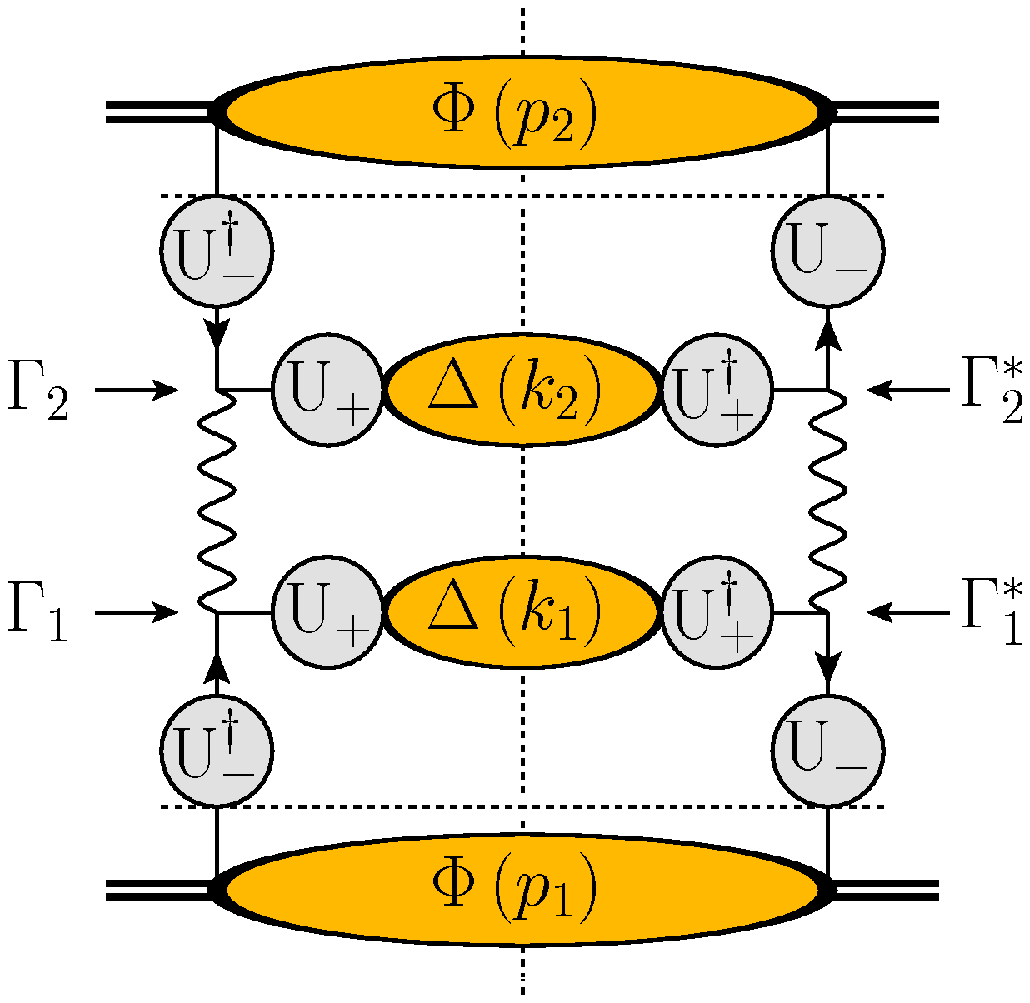,width=0.3\textwidth}
\\[0.2cm]
(a)\hspace{5cm} (b) \hspace{5cm} (c)
\end{center}
\caption{\label{knot}
(a) An example of a diagram with two gluons attaching to the same outgoing
line with momentum $k_2$. 
(b) They result into one gauge connection $U^{[k_2]}_+[p_1,p_2,k_1]$,
which combines all collinear gluons coming from $\Phi(p_1)$, $\Phi(p_2)$
and $\Delta(k_1)$.
(c) Gauge connections appear on all external (colored) lines. }
\end{figure}
\begin{figure}[tb]
\begin{center}
\epsfig{file=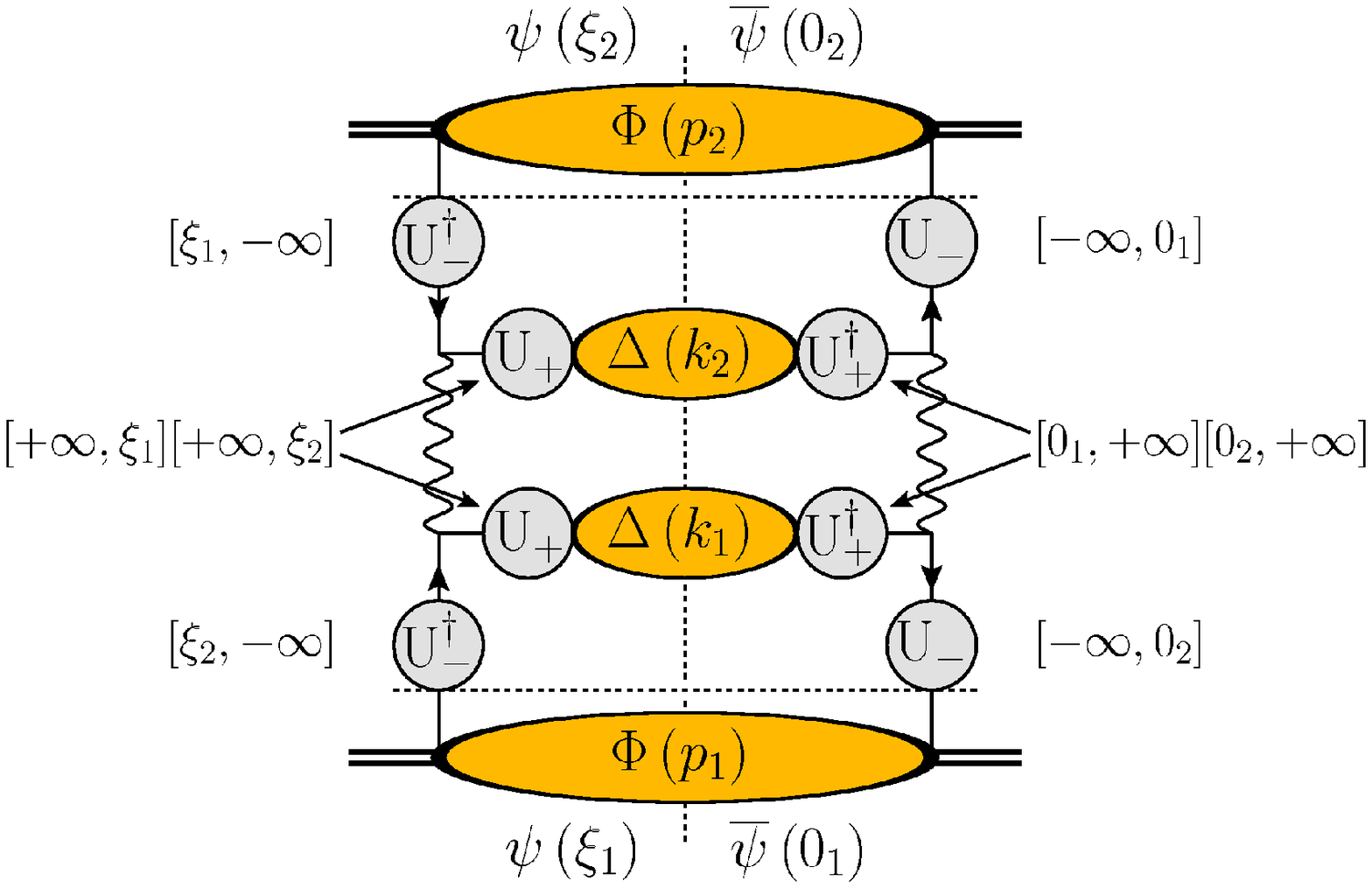,width=0.5\textwidth}
\epsfig{file=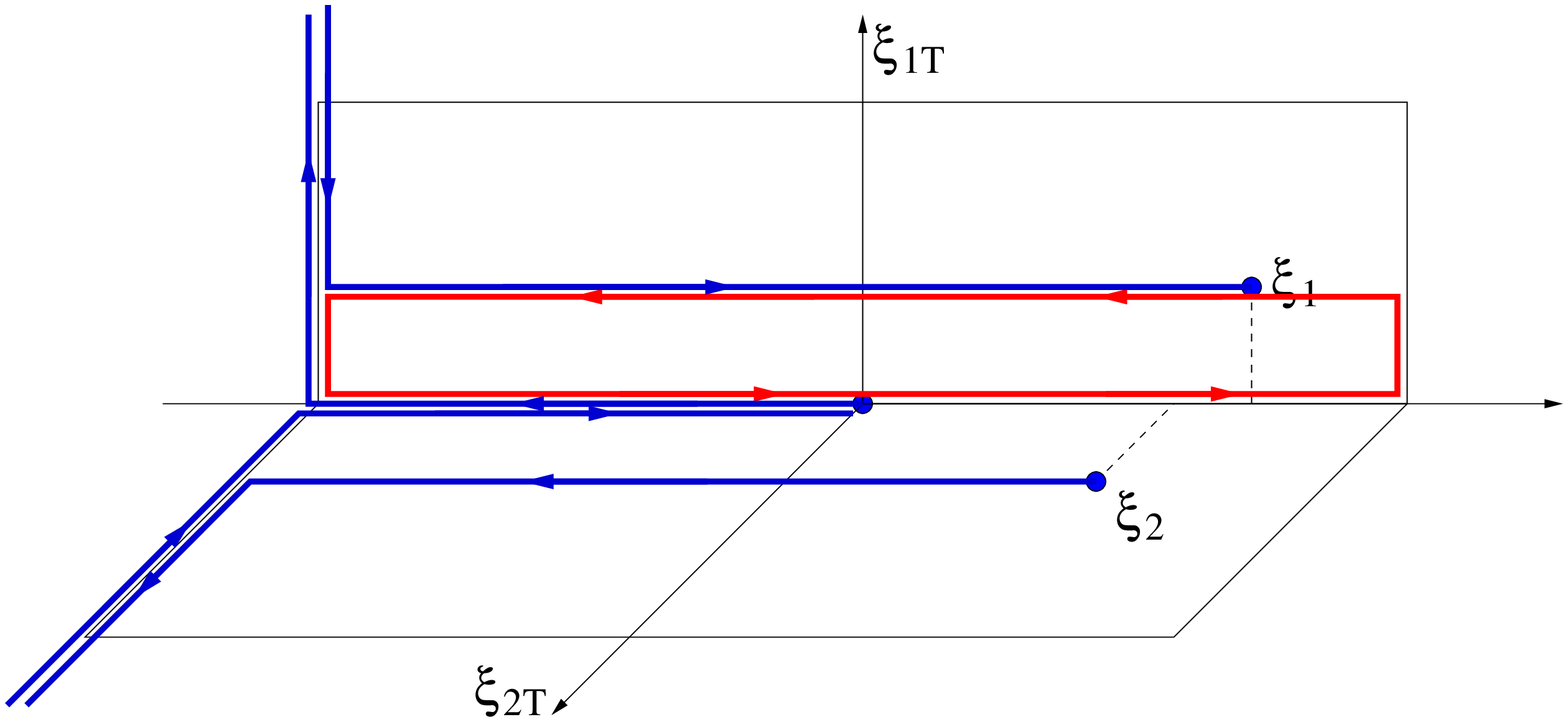,width=0.45\textwidth}
\\[0.2cm]
(a)\hspace{5.5cm} (b)
\end{center}
\caption{\label{cgi-result}
(a) We have indicated for the correlators and gauge connections also the 
actual space-time points they are bridging, limiting ourselves for simplicity
to the coordinates conjugate to $p_1$ (points $0_1$ and $\xi_1$; see also
the discussion following Eq.~\ref{facto2}) 
and $p_2$ (points $0_2$ and $\xi_2$), leaving out
the space-time structure for the fragmentation correlators
for which we would have to include also the coordinates conjugate 
to $k_1$ and $k_2$.
(b) Shown are some combinations of the gauge links including transverse pieces
that can show up in the correlators. These can be Wilson lines and Wilson loops}
\end{figure}

Including gluon insertions from several correlators, for instance those on 
an outgoing quark line coming from two different soft pieces, one from $\Phi(p)$ 
and one from $\Phi(p^\prime)$, such as given in Fig.~\ref{figB1}, gives rise to
intertwined Wilson lines illustrated in Fig.~\ref{knot}. 
Including all multi-gluon interactions as well as the corresponding 
transverse pieces from $\Phi(p_1)$, $\Delta(k_1)$, 
$\Phi(p_2)$ and $\Delta(k_2)$ onto all legs, Eq.~\ref{facto2} generalizes to
\bea
d\sigma & \sim & {\rm Tr}_c\bigl[U_-^{[n]\dagger}[p_2,k_1,k_2]\,\Phi(p_1)
\,U_-^{[n]}[p_2,k_1,k_2]\,\Gamma_1^\ast
\,U_+^{[n]\dagger}[p_1,p_{2},k_{2}]\Delta(k_1)
\,U_+^{[n]}[p_{1},p_{2},k_{2}]\,\Gamma_1\bigr]
\nonumber \\&& \mbox{}\times
{\rm Tr}_c\bigl[U_-^{[n]\dagger}[p_1,k_1,k_2]
\,\Phi(p_2)\,U_-^{[n]}[p_1,k_1,k_2]
\,\Gamma_2^\ast\,U_+^{[n]\dagger}[p_1,p_2,k_1]\,\Delta(k_2)
\,U_+^{[n]}[p_1,p_2,k_1]\,\Gamma_2\bigr],
\label{facto3}
\eea
illustrated in Fig.~\ref{cgi-result}. 
The result for all insertions to a particular leg is a color symmetric
combination of the insertions from all correlators,
\be
U_+^{[n]}[p_1,p_2,k_1]=
{\mathcal S}\{U_+^{[n]}[p_1]U_+^{[n]}[p_2]U_+^{[n]}[k_1]\},
\label{linkbreak1}
\ee
in which the full symmetrization makes the ordering of the three connections 
on the right-hand side irrelevant. 
The resulting expression in Eq.~\ref{facto3} is now color-gauge-invariant. 
The Wilson lines can be taken along a generic $n$-direction, 
but its color structure is fully entangled and it does not allow for
a factorized expression with universal correlators that have their 
own gauge links. Viewing it as a factorized expression it contains
hard amplitudes, soft correlators and gauge connections, where the
gauge connections take care of a `color resetting', which feels all
hadrons that are involved.

If one is interested in an expression for the cross section integrated over 
transverse momenta, one can combine the Wilson lines to and from light-cone 
$\pm \infty$, now all made up of $A^n$ fields, into finite straight-line Wilson lines, 
$U_+^{[n]\dagger}[p_1]\,U_+^{[n]}[p_1]$, since
after the integration over $p_{1\st}$ they not only both run along $n$, 
but they coincide since one also has $0_\st = \xi_\st$. 
Furthermore, it is irrelevant if one started from Wilson lines 
running via plus or via minus infinity, and also the direction $n$ 
is in fact irrelevant, being just the direction of the straight line 
connecting $0$ and $\xi$. 
We recall that the argument $p_1$ or $x_1$, given to the Wilson lines, 
is simply needed to indicate that the fields in that Wilson line belong 
to the correlator 
$\Phi(x_1)$, which is the Fourier transform of the matrix element
$\langle \overline \psi(0)\psi(\xi)\rangle$. 
Thus, in coordinate space one just has the Wilson line in Eq.~\ref{glcol}, 
which connects the points $0$ and $\xi$ in $\Phi(x_1)$ composed of
two pieces. As far as relevant for $\Phi(x_1)$, the Wilson lines
in the first trace form a gauge link, those in
the second trace form a closed loop, which in the collinear situation 
(when $0_\st = \xi_\st$) becomes a unit operator in color space.
One is left with
\bea
\sigma & \sim & {\rm Tr}_c\bigl[U_-^{[n]\dagger}[k_1]\,\Phi(x_1)\,U_-^{[n]}[k_1]
\,\Gamma_1^\ast\,U_-^{[n]\dagger}[p_1]\Delta(z_1)\,U_-^{[n]}[p_1]\bigr]
\nonumber \\&& \mbox{}\times
\,{\rm Tr}_c\bigl[U_-^{[n]\dagger}[k_2]\,\Phi(x_2)\,U_-^{[n]}[k_2]
\,\Gamma_2^\ast\,U_-^{[n]\dagger}[p_2]\,\Delta(z_2)\,U_-^{[n]}[p_2]\,\Gamma_2\bigr].
\eea
The way of turning the gauge connections into gauge links
at the collinear stage is actually just
applying gauge transformations $U^{[n]}_{[a,\xi]}$ (with a
fixed point $a$) to all fields and using the fact that they in the
collinear case only involve fields $A^n$. 
One obtains
\be
\sigma \sim 
\Phi^{[U]}(x_1) \Phi^{[U]}(x_2)
\,\underbrace{\Gamma_1 \,\Gamma_1^\ast \,\Gamma_2 \,\Gamma_2^\ast}_{\hat\Sigma}
\,\Delta^{[U]}(z_1) \,\Delta^{[U]}(z_2),
\label{facto3b}
\ee
where
\bea
\Phi^{[U]}(x) & = & {\rm Tr}_c\bigl[U_\pm^{[n]}[p]\Phi(x)
\,U_\pm^{[n]\dagger}[p]\bigr]
= \left.\int \frac{d\,\xi{\cdot}P}{2\pi}\ e^{i\,p\cdot \xi}
\ \langle P\vert \overline\psi(0)\,U_{[0,\pm\infty]}^{[n]}
\,U_{[\pm\infty,\xi]}^{[n]} \,\psi(\xi)\vert P\rangle
\right|_{LC} 
\nonumber \\ &=&
{\rm Tr}_c\bigl[U^{[n]}[p]\,\Phi(x)\bigr]
= \left.\int \frac{d\,\xi{\cdot}P}{2\pi}\ e^{i\,p\cdot \xi}
\ \langle P\vert \overline\psi(0)\,U^{[n]}_{[0,\xi]}
\,\psi(\xi)\vert P\rangle
\right|_{LC} 
\eea
and
\bea
\Delta^{[U]}(z) & = & \frac{1}{N_c}\,{\rm Tr}_c\bigl[U_\pm^{[n]}[k]\Delta(z)
\,U_\pm^{[n]\dagger}[k]\bigr]
= \frac{1}{N_c}\,{\rm Tr}_c\bigl[\Delta(z)\,U^{[n]\dagger}[k]\bigr]
\nonumber \\ & = &
\left. \int \frac{d\,\xi{\cdot}K_h}{2\pi}\ e^{-ik\cdot \xi}\,
\frac{1}{N_c}\,{\rm Tr}_c \langle 0\vert U_{[\pm\infty,0]}^{[n]}\psi (0) 
a_h^\dagger a_h \overline \psi(\xi)\,U_{[\xi,\pm\infty]}^{[n]} \vert 0 \rangle
\right|_{LC} 
\eea
are the color-gauge-invariant collinear correlators,
including unique gauge links along the light-like separation.
The gauge link being unique, it is usually omitted, writing
$\Phi(x_1)$, $\Phi(x_2)$, $\Delta(z_1)$ and $\Delta(z_2)$.
These (color-gauge-invariant) correlators can be
expanded in terms of the standard unpolarized and polarized parton distribution 
functions and fragmentation functions, respectively. 

Finally, if the transverse momentum in all correlators except one, say $\Phi(p_1)$,
is integrated over, one can shuffle all gauge links into the respective
correlators. We refer to these processes as 1-parton unintegrated 
processes~\cite{Buffing:2011mj}.
All gauge links only involve collinear fields $A^n$ and are
straight-line Wilson lines, except for the
gauge link belonging to $\Phi(p_1)$, for which the transverse separations are 
relevant. In simple processes like SIDIS and DY the longitudinal and transverse
pieces in the remaining TMD correlator then combine into the staple-like gauge 
links in Fig.~\ref{simplelinks}. 
The gauge link for $\Phi(p_1)$ in the case of a process in which the color
flow is more complicated, such as in the example shown in Fig.~\ref{fig2}(b), involves a more complex path, such as 
$U^{[+]}_{[0,\xi]}\,\tr_c(U^{[\Box]})$, in which a staple-like link is
combined with a Wilson loop $U^{\Box}$ = $U^{[+]}_{[0,\xi]}\,U^{[-]}_{[\xi,0]}$ 
(see also Fig.~\ref{cgi-result}(b)). 
The particular TMD correlator $\Phi^{[U]}(x_1,p_{1\st})$ is color-gauge-invariant
and can be expanded in terms of parton distribution functions depending on 
$x$ and $p_\st^2$, although these would in principle still be gauge link-dependent,
for instance for quarks~\cite{Mulders:1995dh,Bacchetta:2006tn},
\begin{eqnarray}
\Phi^{[U]}(x,p_{\st};n)&=&\bigg\{
f^{[U]}_{1}(x,p_\st^2)
-f_{1T}^{\perp[U]}(x,p_\st^2)\,
\frac{\epsilon_{\st}^{\rho\sigma}p_{\st\rho}S_{\st\sigma}}{M}
+g^{[U]}_{1s}(x,p_\st)\gamma_{5}
\nonumber \\&&\mbox{}
+h^{[U]}_{1T}(x,p_\st^2)\,\gamma_5\,\slash S_{\st}
+h_{1s}^{\perp [U]}(x,p_\st)\,\frac{\gamma_5\,\slash p_{\st}}{M}
+ih_{1}^{\perp [U]}(x,p_\st^2)\,\frac{\slash p_{\st}}{M}
\bigg\}\frac{\slash P}{2},
\label{e:parqtmd}
\end{eqnarray}
with the spin vector parametrized as 
$S^\mu = S_{\s L}P^\mu + S^\mu_{\st} + M^2\,S_{\s L}n^\mu$ 
and shorthand notations for $g^{[U]}_{1s}$ and $h_{1s}^{\perp [U]}$,
\begin{eqnarray}
g^{[U]}_{1s}(x,p_T)=S_{\s L} g^{[U]}_{1L}(x,p_{\st}^2)
-\frac{p_{\st}\cdot S_{\st}}{M}g^{[U]}_{1T}(x,p_{\st}^2).
\end{eqnarray}

\section{TMDs of definite rank\label{section3}}

Integrating over components of the parton momenta in the correlators one goes 
from TMDs to collinear correlators and finally to local matrix elements. 
Including moments in these integrations is a way to obtain the coefficients in
an expansion, e.g.\ the way that local matrix elements play a role in the
operator product expansion. The behavior of the local matrix elements, characterized 
by spin and twist, are useful in determining the relevance at leading or 
subleading orders. To study the $x$-dependence 
of the integrated correlator $\Phi(x)$ one constructs the $x^N$ moments. 
To relate these to local matrix elements, the Wilson lines are essential
since by taking moments in $x$ one needs derivatives in $\xi^-$,
\begin{eqnarray}
x^{N}\Phi^{[U]}(x) & = & 
\left. \int \frac{d\,\xi{\cdot} P}{2\pi}\ e^{i\,p{\cdot} \xi}
\ \langle P\vert \overline\psi(0)\,(i\partial^n)^{N}
\,U^n_{[0,\xi]}\,\psi(\xi)\vert P\rangle
\right|_{LC} 
\nonumber \\ & = & 
\left. \int \frac{d\,\xi{\cdot} P}{2\pi}\ e^{i\,p{\cdot} \xi}
\ \langle P\vert \overline\psi(0)\,U^n_{[0,\xi]}\,(iD^n)^{N}
\,\psi(\xi)\vert P\rangle
\right|_{LC}.
\label{momentscollinear}
\end{eqnarray}
Integrating over $x$ one finds the connection of the Mellin moments 
of PDFs involving covariant derivatives $D^n$ as has been already 
discussed following Eq.~\ref{nonlocalfields}.
The local matrix elements have specific anomalous dimensions,
which via an inverse Mellin transform define the splitting functions.

For transverse momentum dependence, we want to expand the TMD correlators as
\be
\Phi(x,p_\st) =\sum_m \Phi^{(m)}(x,p_\st^2)\,p_\st^m (\varphi), 
\ee
where the angle $\varphi$ represents the angular dependence of the transverse vectors 
$p_\st$ and $p_\st^m(\varphi)$ is the symmetric traceless rank $m$ tensor constructed 
from the transverse momenta, i.e.\ 
\be 
p_\st^{\alpha_1\ldots\alpha_m} = p_\st^{\alpha_1}\ldots p_\st^{\alpha_m} - {\rm traces} 
\ \Longleftrightarrow\ p_\st^m(\varphi) 
= \frac{\vert p_\st\vert^m}{2^{m-1}}\,e^{\pm im\varphi}.
\ee
To find the coefficients in such an expansion,
we use transverse moments that involve $p_\st$-weightings of the light-front TMD in Eq.~(\ref{phi-tmd}), including now also a gauge link $U$.
For the simplest gauge links $U^{[\pm]}$, one has
\begin{eqnarray}
p_\st^\alpha\,\Phi^{[\pm]}(x,p_\st;n) & = & 
\int \frac{d\,\xi{\cdot} P\,d^2\xi_\st}{(2\pi)^3}
\ e^{i\,p{\cdot} \xi}
\nonumber\\ && \mbox{}\hspace{-0.1cm} \times 
\langle P\vert \overline\psi(0)\,U^n_{[0,\pm\infty]}
\,U^T_{[0_\st,\xi_\st]}
\,iD_\st^\alpha(\pm\infty)
\,U^n_{[\pm\infty,\xi]}\psi(\xi)\vert P\rangle
\Biggr|_{LF} .
\end{eqnarray}
Integrating over $p_\st$ gives the lowest transverse moment. This moment 
involves twist three (or higher) collinear multi-parton correlators, in particular 
the quark-quark-gluon correlator
\begin{eqnarray}
\Phi^{n\alpha}_{F}(x-x_1,x_1|x) & = &
\int \frac{d\,\xi{\cdot} P\,d\,\eta{\cdot} P}{(2\pi)^2}
\ e^{i\,(p-p_1){\cdot} \xi}\ 
\nonumber\\ && \mbox{}\hspace{0.5cm} \times 
e^{i\,p_1{\cdot} \eta}
\ \langle P\vert \overline\psi(0)\,U^n_{[0,\eta]}\,F^{n\alpha}(\eta)
\,U^n_{[\eta,\xi]}\,\psi(\xi)\vert P\rangle\Biggr|_{LC}.
\end{eqnarray}
In terms of this correlator and the similarly defined correlator 
$\Phi_D^\alpha(x-x_1,x_1|x)$ one finds 
\begin{equation}
\int d^2p_\st\ p_\st^\alpha\,\Phi^{[U]}(x,p_\st)
= \widetilde\Phi_\partial^\alpha(x) + C_G^{[U]}\,\Phi_G^\alpha(x),
\end{equation}
with
\begin{eqnarray}
\widetilde\Phi_\partial^\alpha(x) & = &\Phi_D^\alpha(x) - \Phi_A^\alpha(x)
\nonumber \\ 
&=& \int dx_1\,\Phi_D^\alpha(x-x_1,x_1 | x)
-\int dx_1\,\text{PV}\frac{1}{x_1}\,\Phi_F^{n\alpha}(x-x_1,x_1 | x),
\\
\Phi_G^\alpha(x) &=& \pi\Phi_F^{n\alpha}(x,0 | x).
\end{eqnarray}
The latter is referred to as a gluonic pole or ETQS-matrix 
element~\cite{Efremov:1981sh,Efremov:1984ip,Qiu:1991pp,Qiu:1991wg,Qiu:1998ia,Kanazawa:2000hz}. The function $\Phi_G^\alpha(x)$ is a gluonic 
pole matrix element, corresponding to the emission of a collinear gluon of zero 
momentum~\cite{Brodsky:2002cx,Brodsky:2002rv}. These functions are collinear and independent of the 
gauge link. That dependence is only in the gluonic pole coefficient $C_G^{[U]}$. 
For the simple staple gauge links $U_\pm$ the gluonic pole 
coefficients are $C_G^{[\pm]} = \pm 1$. 
An important property of the two functions showing up in the moments, is their 
behavior under time-reversal. 
While $\widetilde\Phi_\partial^\alpha$ is T-even, $\Phi_G^\alpha$ 
is T-odd. Since time-reversal is a good symmetry of QCD, the appearance of T-even
or T-odd functions in the parametrization of the correlators is linked
to specific observables with this same character. In particular single
spin asymmetries are T-odd observables.

Similarly, we have higher moments,
\be
\Phi_{\partial\partial}^{\alpha\beta\,[U]}(x) = 
\widetilde\Phi_{\partial\partial}^{\alpha\beta}(x) 
+ C_G^{[U]}\,\widetilde\Phi_{\{\partial G\}}^{\alpha\beta}(x)
+ C_{GG,c}^{[U]}\,\Phi_{GG,c}^{\alpha\beta}(x),
\label{e:Phidd}
\ee 
etc. An extra index $c$ is needed if there are multiple possibilities to construct a color singlet as is the case for a field combination $\overline\psi\,GG\,\psi$, namely $\tr_c[GG\psi\overline\psi]$ ($c=1$) and $\tr_c[GG]\,\tr_c[\psi\overline\psi]/N_c$ ($c=2$). The number of gluonic poles determines if we have a T-even or T-odd operator combination. With two color possibilities for a double gluonic pole, there are thus three rank two T-even operator structures, which in the parametrization of the correlator will imply three different pretzelocity functions.
For the staple-like links only one configuration is relevant, having $C_{GG,1}^{[\pm]} = 1$ and $C_{GG,2}^{[\pm]} = 0$. The weighted results also allow a unique parametrization of the gauge link dependent TMD correlators in terms of a finite set of definite rank TMDs depending on $x$ and $p_\st^2$, azimuthal tensors and gluonic pole factors~\cite{Buffing:2012sz,Buffing:2013eka},
\bea
\Phi^{[U]}(x,p_\st) &=& 
\Phi(x,p_\st^2) 
+ \frac{p_{\st i}}{M}\,\widetilde\Phi_\partial^{i}(x,p_\st^2)
+ \frac{p_{\st ij}}{M^2}\,\widetilde\Phi_{\partial\partial}^{ij}(x,p_\st^2)
\nonumber \\ &&\mbox{} +
C_{G}^{[U]}\left\lgroup\frac{p_{\st i}}{M}\,\Phi_{G}^{i}(x,p_\st^2)
+ \frac{p_{\st ij}}{M^2}\,\widetilde\Phi_{\{\partial G\}}^{\,ij}(x,p_\st^2)
\right\rgroup
\nonumber \\ &&\mbox{} +
\sum_c C_{GG,c}^{[U]}\frac{p_{\st ij}}{M^2}\,\Gamma_{GG,c}^{ij}(x,p_\st^2).
\label{e:TMDstructure}
\eea
Depending on partons (quarks or gluons) and target, there is a maximum rank, which 
for quarks in a nucleon is rank 2. For gluons in a nucleon one has to go up to rank 3. 
Actually for the highest rank, time-reversal symmetry does not allow a time-reversal odd 
rank 2 correlator, i.e.\ $\widetilde\Phi_{\{\partial G\}} = 0$. Note that since the 
tensors $p_\st^{ij}$ on the rhs of Eq.~\ref{e:TMDstructure} are traceless and symmetric, 
the correlators they multiply also must be made traceless in order to make the 
identification of the correlators unique.

The situation with universality for fragmentation functions is easier because
the gluonic pole matrix elements vanish in that 
case~\cite{Collins:2004nx,Gamberg:2008yt,Meissner:2008yf,Gamberg:2010gp}. Nevertheless, there 
exist T-odd fragmentation functions, but their QCD operator structure 
is T-even. These T-odd functions then appear in the parametrization
of $\widetilde\Phi_\partial^\alpha$.
Hence, there is no process dependence from gluonic pole factors for fragmentation functions.

The reproduction of the transverse moments provides the proper
identification of universal TMD functions, e.g.\ for quarks
\bea
&&\Phi(x,p_\st^2) = 
\bigg\{
f_{1}(x,p_\st^2)
+S_{\s L}\,g_{1}(x,p_\st^2)\gamma_{5}
+h_{1}(x,p_\st^2)\gamma_5\,\slash S_{\st}
\bigg\}\frac{\slash P}{2},
\label{rank0}
\\
&&\frac{p_{\st i}}{M}\,\widetilde\Phi_\partial^{i}(x,p_\st^2) =
\bigg\{
h_{1L}^{\perp}(x,p_\st^2)\,S_{\s L}\frac{\gamma_5\,\slash p_{\st}}{M}
-g_{1T}(x,p_\st^2)\,\frac{p_\st{\cdot}S_\st}{M}\gamma_{5}
\bigg\}\frac{\slash P}{2},
\\ &&
\frac{p_{\st i}}{M}\,\Phi_{G}^{i}(x,p_\st^2)
= \bigg\{
-f_{1T}^{\perp}(x,p_\st^2)\,
\frac{\epsilon_{\st}^{\rho\sigma}p_{\st\rho}S_{\st\sigma}}{M}
+ih_{1}^{\perp}(x,p_\st^2)\,\frac{\slash p_{\st}}{M}
\bigg\}\frac{\slash P}{2},
\\
&&
\frac{p_{\st ij}}{M^2}\widetilde\Phi_{\partial\partial}^{ij}(x,p_\st^2)
= h_{1T}^{\perp (A)}(x,p_\st^2)
\,\frac{p_{\st ij}S_\st^i\,\gamma_5\gamma_{\st}^j}{M^2}
\,\frac{\slash P}{2},
\\ &&
\frac{p_{\st ij}}{M^2}\Phi_{GG,1}^{ij}(x,p_\st^2)
= h_{1T}^{\perp (B1)}(x,p_\st^2)
\,\frac{p_{\st ij}S_\st^i\,\gamma_5\gamma_{\st}^j}{M^2}
\,\frac{\slash P}{2},
\\ &&
\frac{p_{\st ij}}{M^2}\Phi_{GG,2}^{ij}(x,p_\st^2)
= h_{1T}^{\perp (B2)}(x,p_\st^2)
\,\frac{p_{\st ij}S_\st^i\,\gamma_5\gamma_{\st}^j}{M^2}
\,\frac{\slash P}{2},
\\ &&
\frac{p_{\st ij}}{M^2}\widetilde\Phi_{\{\partial G\}}^{ij}(x,p_\st^2)
= 0. \label{PhipartialG}
\eea
We note that the rank zero functions in Eq.~(\ref{rank0}) depend on
$x$ and $p_\st^2$ and involve traces,
to be precise
$g_1(x,p_\st^2) = g_{1L}^{[U]}(x,p_\st^2)$ and
$h_1(x,p_\st^2) = h_{1T}^{[U]}(x,p_\st^2) 
- (p_\st^2/2M^2)\,h_{1T}^{\perp [U]}(x,p_\st^2)$. 
As remarked before, for the pretzelocity there are three
universal functions with in general
\be 
h_{1T}^{\perp [U]}(x,p_\st^2)
= h_{1T}^{\perp (A)}(x,p_\st^2)
+ C_{GG,1}^{[U]}\,h_{1T}^{\perp (B1)}(x,p_\st^2)
+ C_{GG,2}^{[U]}\,h_{1T}^{\perp (B2)}(x,p_\st^2).
\ee
For the simplest gauge links we have $C_{GG,1}^{[\pm]} = 1$ and
$C_{GG,2}^{[\pm]} = 0$, which shows e.g.\ that 
$h_{1T}^{\perp [{\rm SIDIS}]}(x,p_\st^2)
=h_{1T}^{\perp [{\rm DY}]}(x,p_\st^2)$, but that for other
processes (with more complicated gauge links) other
combinations of the three possible pretzelocity functions occur.
For a spin 1/2 target the above set of TMDs is complete. There are
no higher rank functions. For a spin 1 target and for gluons, there
are higher rank functions~\cite{Buffing:2012sz,Buffing:2013eka,Buffing:2013kca}.
For the fragmentation correlator there is for rank 2 only a single 
(T-even) pretzelocity function 
$H_{1T}^\perp(z,k_\st^2)$ appearing in the parametrization of the 
correlator $\Delta_{\partial\partial}^{\alpha\beta}(x,p_\st^2)$.

\section{Conclusions}
\label{sec:conclusions}

We have discussed color gauge invariance for TMD correlators. These involve
parts along the light-cone and transverse pieces off the light-cone. If the
gauge link in a single hadron correlator is considered, one can construct TMDs of definite rank leading to an expansion as in Eq.~(\ref{e:TMDstructure}). In this decomposition, we have made an expansion of the quark correlator into irreducible tensors multiplying correlators containing operator combinations of gluons, covariant derivatives and $A$-fields, the latter in the combination
$i\partial = iD - gA$. In the decomposition gluonic pole factors
contain the gauge link dependence, which are calculated from
the transverse moments. 
The correlators of definite rank in turn are parametrized in terms
of the universal TMD PDFs depending on $x$ and $p_\st^2$, such as given 
by Eqs~(\ref{rank0})-(\ref{PhipartialG}). The process dependence for
a particular TMD PDF is in the same gluonic pole factors that appear 
in the expansion in Eq.~(\ref{e:TMDstructure}).

An analysis for a quark spin 1/2 target shows that the process 
dependence is not strictly confined to the T-odd functions, such as the 
Sivers or the Boer-Mulders functions. In fact, there exist three T-even 
pretzelocity functions. For fragmentation the 
TMD PFFs are already universal since gluonic pole matrix elements
vanish for fragmentation correlators.
Quark TMDs can also be studied for higher spins or gluon TMD PDFs. 
While for a spin 1/2 target one 
has at most rank two TMDs, one has for higher spins and gluon TMDs
also higher rank functions, while also the color and gauge link
structure is richer. To study the appearance of the TMDs in cross sections,
in particular in situations in which the transverse partonic momenta of
several hadron correlators are involved, requires care~\cite{Buffing:2013dxa}.
The knowledge of the operator structure including its rank most likely will
also be relevant in the detailed study of the QCD evolution of the full
set of TMDs~\cite{Collins:2011zzd,Collins:2011ca,Collins:2008sg,Aybat:2011zv,Aybat:2011ge,Aybat:2011ta,Rogers:2013zha}.

\begin{acknowledgements}
This research is part of the research program of the ``Stichting voor Fundamenteel Onderzoek der Materie (FOM)'', which is financially supported by the ``Nederlandse Organisatie voor Wetenschappelijk Onderzoek (NWO)''. Part of the research is supported by the FP7 EU-programmes HadronPhysics3 (contract no 283286) and the ERC Advanced Grant program QWORK (contract no 320389). We acknowledge discussions with colleagues on 
some of this work, among them Dani\"el Boer, Wilco den Dunnen and Asmita Mukherjee.
\end{acknowledgements}




\end{document}